\documentclass[twocolumn,preprintnumbers,amsmath,amssymb]{revtex4-1}
\usepackage{graphicx}
\usepackage{color}
\usepackage{caption}
\usepackage{subfigure}
\begin{document}

\title{
Spatiotemporal intermittency and localized dynamic fluctuations upon approaching the glass transition}

\author{J. Ariel Rodriguez Fris$^1$}
\author{Eric R. Weeks$^2$}
\author{Francesco Sciortino$^{3,4}$}
\author{Gustavo A. Appignanesi$^1$}

 \affiliation{
$^1$ INQUISUR, Departamento de Qu\'{i}mica, Universidad Nacional del Sur (UNS)-CONICET, Avenida Alem 1253, 8000 Bah\'{i}a Blanca, Argentina.\\
$^2$Physics Department, Emory University, Atlanta, Georgia 30322, USA.\\
$^3$Dipartimento di Fisica,  Sapienza Universita' di Roma, 
Piazzale A. Moro 5, Roma 00185, Italy.\\
$^4$CNR-ISC, c/o Sapienza, Piazzale A. Moro
5, Roma 00185, Italy.\\
}

\date{\today}

\begin{abstract}
We introduce a new and robust method for characterizing spatially and temporally heterogeneous behavior within a system based on the evolution of dynamic fuctuations once averaged over different space lengths and time scales.
We  apply it to investigate the dynamics  in two canonical systems as the glass transition is approached: a simulated Lennard-Jones glass-former and a real dense colloidal suspensions. 
We find that in both cases the onset of glassines is marked by spatially localized dynamic fluctuations originating in  regions of correlated mobile particles. By removing the trivial system size dependence of the fluctuations  we show that such regions  contain  tens to hundreds of particles for time scales corresponding to maximally non-Gaussian dynamics.
\end{abstract}


\maketitle

\section{Introduction}

Glasses are solid materials with disordered liquid-like
structure.  These are typically formed by rapidly quenching
a liquid from a hot  to a cold temperature,
or compressing a liquid from a low  to a high
pressure \cite{angell95,angell00}.  How the
transition from an equilibrium liquid to an out of equilibrium
glass takes places is  highly debated, with many different and contrasting  interpretations proposed; see Refs.~\cite{langer14,chandler10,biroli13,ediger12,lubchenko07,cavagna09}
for reviews.  One point is known:  the onset of unusual behavior
within a sample precedes the glass transition. ``Supercooled
liquids'', despite their metastable equilibrium, have a markedly higher viscosity $\eta$ than normal liquids.  This step rise of $\eta$  is associate to the onset of  dynamical heterogeneity:  diffusive motion  takes
place in a spatially and temporally heterogeneous fashion
\cite{sillescu99,ediger00,glotzer00,hempel00,richert02}.  At
any given time, some regions within the sample are frozen, while
other regions are quite mobile.  The mobile regions are
characterized by ``cooperative'' motion where localized groups of molecules
have nearly simultaneous large displacements
\cite{glotzer00,kob97,donati98,donati99}.  Over time, the
locations of faster and slower dynamics change, such that at any
given position the dynamics are temporally heterogeneous as well
\cite{chandler10,garrahan03,keys11}.

In the last two decades, a variety of methods
have been proposed and implemented to characterize such dynamical heterogeneities.  Early work studied
simulations of soft particles or Lennard-Jones particles and
identified subsets of particles that had large displacements $\Delta r$
\cite{hurley95,hurley96,kob97,donati98,donati99}, showing that
these formed spatially localized clusters.  A key result was the identification of the non-Gaussian time scale
$\Delta t^*$ as an important time scale related to these clusters
\cite{hurley96,kob97,marcus99}.  This time scale is identified by
examining the behavior of the non-Gaussian parameter
$\alpha_2(\Delta t) = 3 \langle \Delta r^4 \rangle / (5 \langle
\Delta r^2\rangle^2) - 1$, a quantity derived from the moments of
the displacement distribution $P(\Delta r)$ \cite{rahman64}.
$\Delta t^*$ is the time scale for
which $\alpha_2(\Delta t)$ is maximal.
$\alpha_2$ is zero when $P(\Delta r)$ is a Gaussian, which holds to a good approximation for simple liquids.  For dynamically heterogeneous
supercooled liquids, $\alpha_2 > 0$ indicating that particles
with large displacements occur more frequently than would occur
for a Gaussian distribution.    Identifying the particles
responsible for the large $\alpha_2$ value is expected to provide
information on the cooperatively moving clusters~\cite{kob97,donati98,donati99} and their structural and dynamic properties.   Early
studies~\cite{kob97,donati98,donati99} used various somewhat arbitrary criteria to define mobile
particles.  Later work examined spatial correlation functions
averaged over all particles in various ways attempting to
identify the length
and time scales of dynamical heterogeneity 
\cite{donati99,flenner07,doliwa00,weeks07cor,glotzer00,lacevic03,keys07,appignanesi06,appignanesi09}.

In this manuscript, we present a new and robust analysis method to
characterize spatial and temporal dynamical heterogeneity 
that does not requires any a priori definition of  
particle mobility.  
In particular, we here use the system mean square displacement as a ``null
hypothesis'' for particle motion, and quantify spatially and
temporally localized deviations of particle motion away from this
null hypothesis.  We apply this method to the Kob-Andersen Lennard-Jones
glassforming system~\cite{kob95a} and to colloidal supercooled
liquid data~\cite{weeks00}.  Our results show that dynamical
heterogeneity is most obvious for subsystems comprised of tens to
hundreds of particles, with the size growing as the glass
transition is approached.  Additionally, we examine how dynamical
heterogeneity becomes averaged out at larger length scales.
As a byproduct we confirm that $\Delta t^*$
is the time scale of maximum heterogeneity.  While our method of localized fluctuations is applied
to particle displacements, the idea is generalizable to other
quantities which may have spatiotemporal fluctuations such as 
structure~\cite{royall15,royall08}.  An advantage of our technique
is that it is applicable to small data sets such as the
experimental colloidal data we use.

\section{Simulation and Experimental Details}

\subsection{Simulation }

We use LAMMPS to simulate the Kob-Andersen binary Lennard-Jones
glassforming system \cite{kob95a}.  Briefly, this is an 80:20
mixture of $A$ and $B$ particles.  The particles interact via the
Lennard-Jones potential \cite{lennardjones24}
\begin{equation}
U_{\alpha \beta}(r) = 4\epsilon_{\alpha \beta} \left[ 
\left(\frac{\sigma_{\alpha \beta}}{r}\right)^{12} -
\left(\frac{\sigma_{\alpha \beta}}{r}\right)^{6} \right]
\end{equation}
with $\alpha,\beta \in {A,B}$.  $A$ and $B$ particles have the same
mass.  The energy scales are
$\epsilon_{AA} = 1.0$, $\epsilon_{AB}=1.5$, and $\epsilon{BB}=0.5$.  The
size scales are $\sigma_{AA} = 1.0$, $\sigma_{AB} = 0.8$, and
$\sigma_{BB} = 0.88$, chosen so that $A$ and $B$ particles are
encouraged to mix rather than segregate, and thus crystallization
is frustrated \cite{kob95a}.  For most of our results we present
data with 8000 total particles, and for our analysis we consider
the $N=6400$ $A$ particles. To verify that our analysis is not
biased by finite size,  in a few cases we compare
with a $N=8\times 10^5$ data set.  Periodic boundary conditions were used
with a cubical box.

\subsection{Experiment}

Colloids have long been used as
model systems to study the glass transition
\cite{hunter12rpp,marshall90,pusey87,pusey86,hartl01,vanmegen91,bartsch93,meller92,lindsay82}.
We reanalyze previously published data from experiments
using confocal microscopy to observe dense colloidal samples
\cite{weeks00}.  The samples were sterically stabilized colloidal
poly-(methyl methacrylate) for which the key control parameter
is the volume fraction $\phi$ \cite{hunter12rpp}.  The glass
transition for this experiment occurred at $\phi_g = 0.58$.  Here we
examine data with $\phi = 0.46, 0.52, 0.56$ with $N\approx 2500,
2900, 3100$ particles respectively in the observation volume.
The particles were a single species with mean particle diameter
2.36~$\mu$m and a polydispersity of 0.045 \cite{kurita12}
and were slightly charged.  Confocal microscopy and particle
tracking was used to follow the positions of the particles
in three dimensions \cite{dinsmore01,crocker96}. The imaging
volume was rectangular with an aspect ratio roughly $5:5:1$;
see Ref.~\cite{weeks00} for details.

\section{Results}

We aim at characterizing the growing spatial and temporal fluctuations on approaching the glass transition without introducing any  arbitrary cut-off quantity.
Following prior work~\cite{ohmine1993fluctuation,appignanesi06,appignanesi09}, we start
by defining the distance matrix $\Delta^2(t',t'')$, an object
that represents the average of the  squared particle displacements
between time $t'$ and $t''$ of a collection of the $N$ particles  belonging to a predefined set $S$ ($S$
may be the entire system or some subsystem):
\begin{eqnarray}
\Delta^2(t',t'') &\equiv & \frac{1}{N} \sum_{i=1}^{N}
| \vec{r}_i(t') - \vec{r}_i(t'')|^2\\
&=& \langle | \vec{r}_i(t') - \vec{r}_i(t'')|^2 \rangle_{i \in S}
\end{eqnarray}
where the angle brackets indicate an average over the $N$
particles in $S$.  
Further averaging 
 $\Delta^2(t',t'')$ over all pairs $t'$ and $t''$ such that
$t''-t'=\Delta t$  produces the well-known average mean square displacement $M^2(\Delta t)$ of particles in $S$. More precisely
\begin{equation}
M^2(\Delta t) = \langle \Delta^2(t',t'') \rangle_{t''-t'=\Delta t}
\end{equation}
where the average is over $t',t''$ with fixed time interval
$\Delta t=|t''-t'|$ and also over all of the particles in $S$. 
Assuming stationary dynamics (true as long as
the system is not aging), for a sufficiently large $\Delta t$, 
$\lim_{\Delta t \rightarrow \infty}\Delta^2(t',t'+\Delta t) = M^2(\Delta t)$

\begin{figure*}[hbtp]
{
\begin{center}
\mbox{
 \leavevmode
 \subfigure   {\includegraphics[width=0.25\textwidth,angle=0, clip=true]{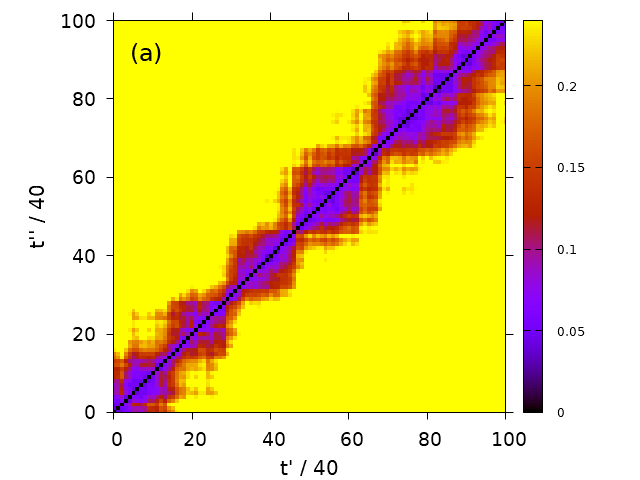}}
   \leavevmode
    \subfigure 
{\includegraphics[width=0.25\textwidth,angle=0, clip=true]{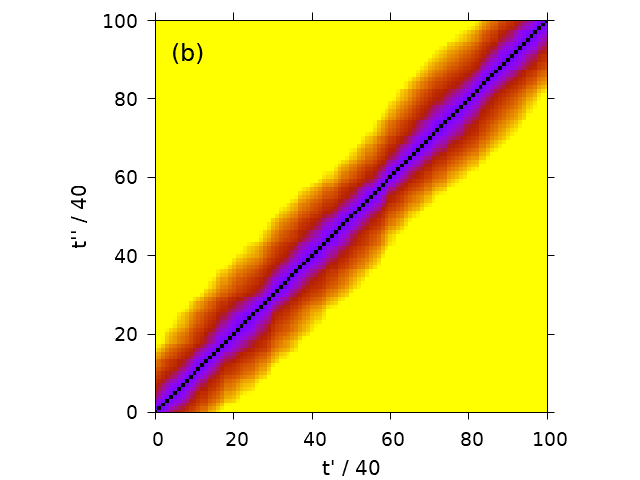}}
 \leavevmode
     \subfigure
{\includegraphics[width=0.25\textwidth,angle=0, clip=true]{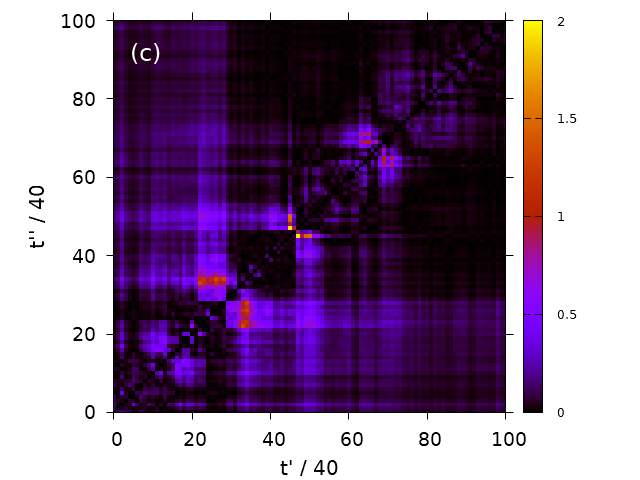}}
 \leavevmode
     \subfigure
{\includegraphics[width=0.2\textwidth,angle=0, clip=true]{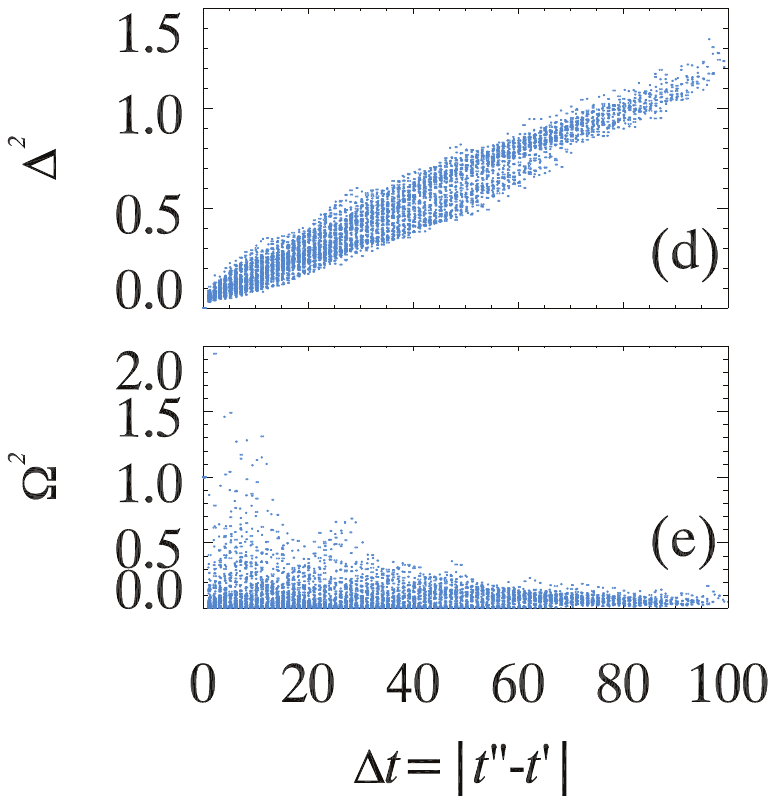}}
}
\end{center}
}
\caption{(a) Contour plot of the distance matrix
$\Delta^2(t',t'')$ for a binary Lennard-Jones system at $T=0.50$
within a cubical subsystem containing 125 particles.  The legend
indicates the values of the gray scale.  (b) Contour plot of the
same system for the full 8000 particle simulation.  Images taken
from Ref.~\cite{appignanesi09}, permission pending.  (c)
$\Omega(t',t'')$.
The data correspond to the same subsystem as in (a).
In the image shown, darker points indicate larger values.  
(d) The values of $\Delta^2$  
as a function of $\Delta t = |t' - t''|$ for the data shown in (a).  
For small $\Delta t$ this
tends to 0, and for intermediate $\Delta t$ the scatter indicates
the temporal heterogeneity in this subsystem.  The overall
increase mirrors the mean square displacement $M^2(\Delta t)$.
(e) $\Omega^2(\Delta t)$ for the data shown in (c).  The scatter
at small $\Delta t$ reflects 
a function of $\Delta t$, which has much larger fluctuations at
short $\Delta t$, reflecting non-Gaussian statistics of the
displacements at those time scales.
} \label{dm}
\end{figure*}

For small systems under glassy relaxation conditions, $\Delta^2$ has temporal fluctuations,
as shown in Fig.~\ref{dm}(a) for a sub-system of $N=125$ particles.   Darker regions indicate time intervals $(t',t'')$ over which this subsystem has relatively
little particle motion.  Clearly, there are specific times for
which this subsystem undergoes fairly large changes, signalled by
larger displacements and substantially different particle
positions.   It is expected~\cite{la2004static} than on increasing the
sample size well beyond any dynamic correlation length, different  regions
of the system will independently display such burst motion,  such that  $\Delta^2$  for larger systems will appear much smoother, as shown in
Fig.~\ref{dm}(b) for the fully system of $N=8000$ particles.  As for the long time limit,  for a sufficiently large system $S$, 
$\lim_{N \rightarrow \infty}\Delta^2(t',t'+\Delta t) = M^2(\Delta t)$

The question we turn to is how the large system limit is
reached. In particular, we wish to use the approach to the
large system limit to characterize the spatial scale of dynamical
heterogeneities.  The obvious features of Fig.~\ref{dm}(a) are the
large fluctuations that differentiate it from Fig.~\ref{dm}(b).
This motivates us to consider the normalized difference between
$\Delta^2$ and the expectation for a large system, defined by
\begin{equation}
\Omega_S^2(t',t'') = 
\frac{ [\Delta^2(t',t'') -  M^2(\Delta t)]^2} {[ M^2(\Delta t)]^2}
\end{equation}
with the convention $\Delta t = |t'' - t'|$.  $\Omega_S^2$,
a measure of the  dynamic intermittency, 
represents  the matrix of normalized squared deviations from
the mean value for the particles squared displacements and will
be equal to zero when $\Delta^2$ is calculated for sufficiently
large systems, for which time averages and space averages are
equivalent and $\Delta^2 = M^2$.  Otherwise, $\Omega_S^2 > 0$
and larger values indicate larger deviations between $\Delta^2$
(local in both space and time) and the expectation for a large system
(that is, $M^2$, a quantity averaged over all space and all time).

An example of $\Omega^2_S(t',t'')$ is shown in Fig.~\ref{dm}(c),
where the darker regions indicate time periods for which the
mean motion within the sub-volume is anomalously larger or smaller
compared to the expectation from  $M^2$.  
Fig.~\ref{dm}(d) and \ref{dm}(e) show scatter
plots of the values of $\Delta^2$ and $\Omega_S^2$ as functions
of $t''-t'$, taken from the data of Fig.~\ref{dm}(a) and
\ref{dm}(c) respectively.  $\Delta^2$ starts at the origin and
rises, consistent with the idea that on average it should behave
similar to the mean square displacement $M^2(\Delta t)$.  At
intermediate time scales the scatter in the data of
Fig.~\ref{dm}(d) indicates the temporal fluctuations of the
motion.  In contrast, $\Omega_S^2$ in Fig.~\ref{dm}(e) has large
fluctuations at the shortest time scales, indicating large
fluctuations of the motion relative to $M^2(\Delta t)$ on those
time scales.  At larger time scales, the temporal averaging
reduces $\Omega_S^2$ toward zero.

As defined, $\Omega^2_S$ is local in space and time. 
To focus on  the space dependence of the fluctuations 
we need to evaluate a single nondimensional scalar quantity $\Omega(N)$
characterizing the mobility fluctuations for subsystems
of size $N$ (the ratio of the dispersion to the average~\cite{chandlerbook}  for the particle squared displacements).
To do so we  partition the system into distinct cubical boxes containing $N$
particles each and evaluate  the sum of 
$\Omega_S^2(t',t'')$ over all time pairs ($t'$, $t''$)  [i.e. the sum over all
points entering in the scatter-plot as the one shown in Fig.~\ref{dm}(e)] 
divided by the number of such pairs  for each of the boxes.  We then 
average the resulting number over all boxes 
 and finally take the square root of the result.  
This procedure
yields the desired scalar quantity $\Omega(N)$.  Note that the specific 
values of $\Omega(N)$ will
depend on the total time studied, that is, the maximum of $|t'' -
t'|$ that is studied. 
 As is apparent from Fig.~\ref{dm}(c) and
\ref{dm}(e), at large $|t''-t'|$, $\Omega_S$ decays to zero, and
the more of this included in the average, the smaller $\Omega(N)$
will be.  However, for a given data set, what will matter is the
$N$-dependence which is insensitive to the total time studied, as
long as that time is sufficient to capture the temporal
fluctuations seen in Fig.~\ref{dm}(c).  In practice, we ensure
that our data sets have a total duration of $\approx 10 \Delta t^*$
where $\Delta t^*$ depends on the temperature (for the Lennard-Jones
simulations) or the volume fraction (for the colloidal
experiments).  This then will allow for a sensible comparison
between different data sets.

\begin{figure*}[hbtp]
{
\begin{center}
\mbox{
 \leavevmode
 \subfigure{\includegraphics[width=0.27\textwidth,angle=0, clip=true]{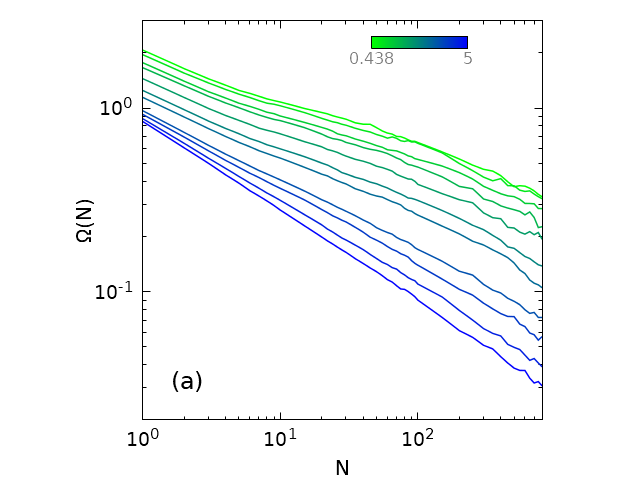}}
\hskip -.8 cm     
   \leavevmode
    \subfigure
{\includegraphics[width=0.273\textwidth,angle=0, clip=true]{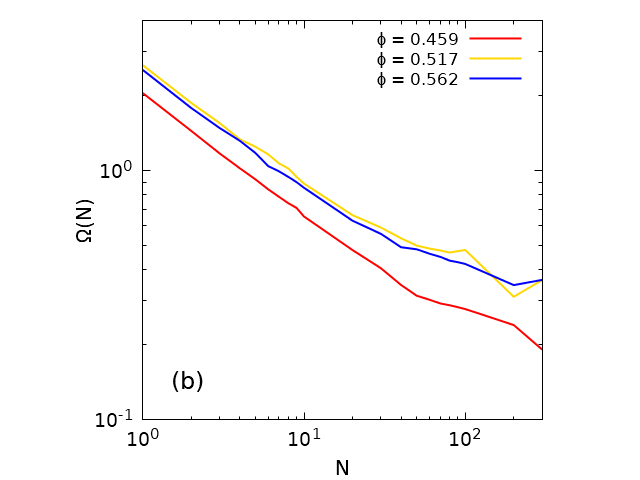}}
\hskip -0.8 cm     
\leavevmode
     \subfigure
{\includegraphics[width=0.27\textwidth,angle=0, clip=true]{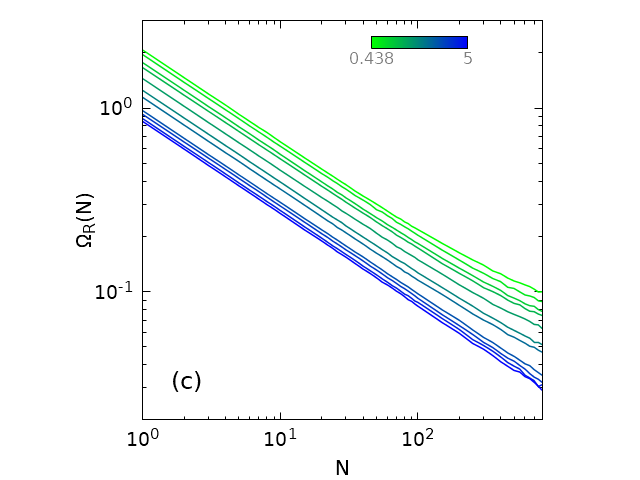}}
 \hskip -0.8 cm     
 \leavevmode
     \subfigure 
{\includegraphics[width=0.27\textwidth,angle=0, clip=true]{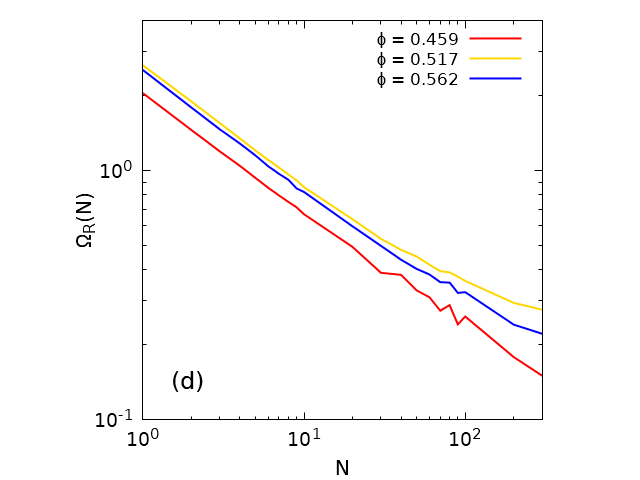}}
}
\end{center}
}
\vskip -0.7 cm
{\caption{$\Omega({\rm N})$ and $\Omega_R({\rm N})$
as a function of subsystem size $N$ for the binary Lennard-Jones system [(a) and (c)] for
different temperatures as indicated and for 
the experimental colloidal suspension  [(b) and (d)] at different volume fractions $\phi$ as indicated.  In the case of $\Omega({\rm N})$ the $N$ particles 
are part of the same subsystem. In the case of  $\Omega_R({\rm N})$
 the $N$ particles are 
selected {\it randomly} among all particles in the system.
\label{omega}}}
\end{figure*}

$\Omega(N)$ for the Lennard-Jones system is plotted in
Fig.~\ref{omega}(a), and for the colloidal experiments in
Fig.~\ref{omega}(b).  In both cases, we see how the dynamical
fluctuations average out for larger subsystem sizes.
Notably, the systems closer to the glass transition
require larger subsystems before the dynamical fluctuations are
averaged out -- colder systems for the LJ data (a), and higher volume
fraction systems for the colloidal data (b).

\begin{figure}[h]
\begin{center}
\mbox{
	\leavevmode
     \subfigure 
		{\includegraphics[width=4.8cm]{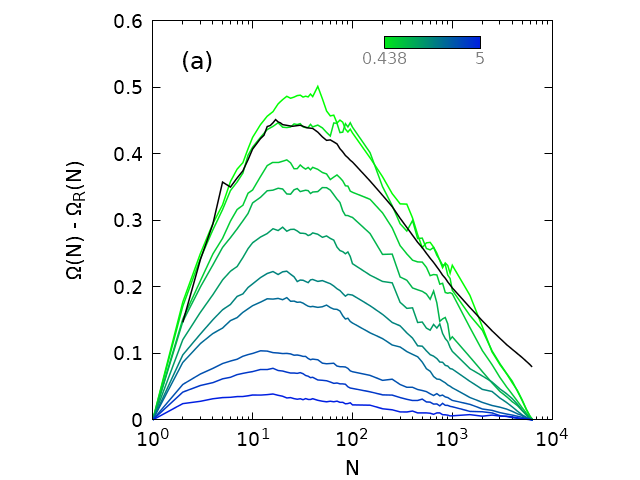}}
\hskip -0.7 cm			
		\leavevmode
     \subfigure
				{\includegraphics[width=4.8cm]{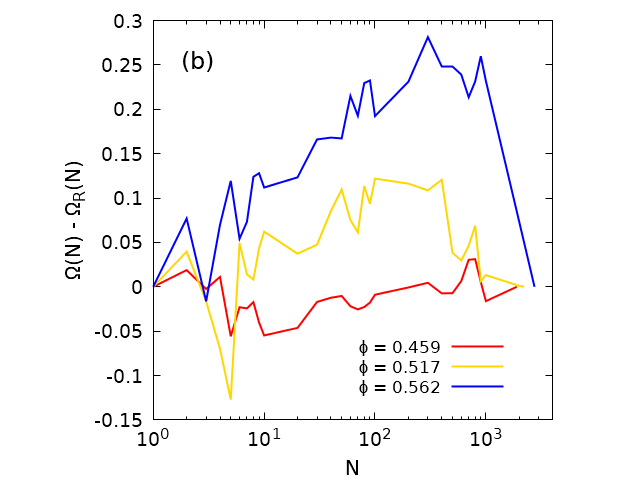}}
				}
	\end{center}
	\vskip -0.7 cm			

\caption{Subtracting $\Omega$ calculated for randomly chosen
particles from $\Omega$ calculated for compact subsystems. (a)
For the LJ system; (b) for the colloidal suspensions. 
The black curve in (a)  shows the same quantity evaluate for a system of 
 1M particles at $T=0.466$, showing that the position of the maximum 
 is not affected by the finite size of the investigated system.
}
\label{subtract}
\end{figure}

An interesting point emerges when analyzing the functional
form of the decay of $\Omega(N)$ with $N$. From
the high temperature data of
Fig.~\ref{omega}(a) we observe
the spatially
localized dynamic fluctuations measured by $\Omega$ display
the usual $N^{-1/2}$ size scaling dependence.  This reflects that
particle motion is nearly spatially uncorrelated within a
subsystem and so the average of $\Delta^2(N)$ converges to the
large-system limit $M^2$ as $N^{-1/2}$.
However, for glassier systems
a clear departure from this trivial behavior
is seen, and  the decay of $\Omega(N)$ 
gets slower.  The fact that the localized dynamical
fluctuations are higher than expected, persisting at large system
sizes, speaks of the existence of regions of correlated mobile
particles, an effect that is more pronounced upon supercooling
\cite{glotzer00,kob97,donati98,donati99,weeks00,weeks07cor}.
We truncate our calculation at 
$N=N_{max}/10$, as we desire at 
least 10 subsystems to evaluate a reasonable
$\Omega(N)$.
We note that both for the
Lennard-Jones system and the colloidal suspensions, the behavior
of the curves of Fig.~\ref{omega} at very large system sizes is due
to lack of statistics (the subsystems are not small as compared to
the large system and, thus, we can only average over a few of them).

As evident from Fig.~\ref{omega}, for the smallest possible
subsystems ($N=1$), the increasing value of $\Omega$ as the glass
transition is approached reflects the  well-known  increasing non-Gaussian nature of the
displacement distribution~\cite{kob95a}.  In fact this points out
a limitation of $\Omega$, in that large values of $\Omega$ can
reflect either spatial fluctuations in the dynamics or simply a
non-Gaussian distribution of displacements.  To remove the latter
influence (that is, to remove the trivial system size dependence
and thus to highlight the local correlations) we separately compute
$\Omega_R(N)$ based not on compact subsystems of size $N$ but on
$N$ randomly chosen particles.  Here the subscript $R$ indicates
an average over many such randomly chosen subsets.  For the
Lennard-Jones system, this is plotted in Fig.~\ref{omega}(c),
showing different behavior from Fig.~\ref{omega}(a) for the
colder temperature data. In fact, now the randomly distributed
dynamical fluctuations quantified by $\Omega_R(S)$ display the
typical $N^{-1/2}$ decay at all temperatures. Similar behavior
is found in Fig.~\ref{omega}(d) for the colloidal suspensions,
as compared to Fig.~\ref{omega}(b). Again, the $N^{-1/2}$ scaling
is recovered at all volume fractions.

\begin{figure}[th]
\begin{center}
\mbox{
	\leavevmode
     \subfigure 
		{\includegraphics[width=4.8cm]{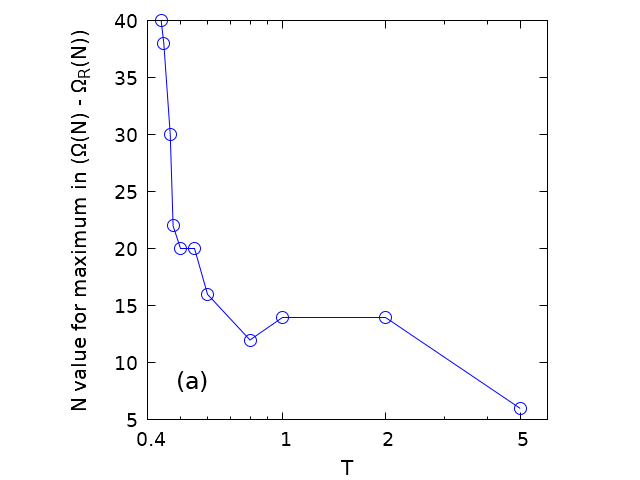}}
\hskip -0.8 cm		
		\leavevmode
     \subfigure 
				{\includegraphics[width=4.8cm]{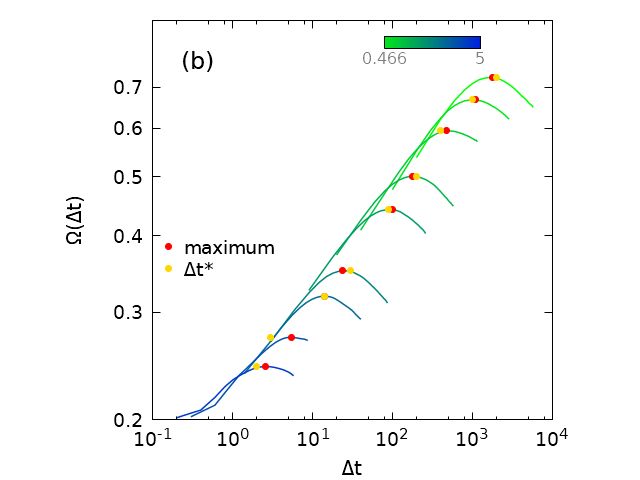}}
				}
	\end{center}
	\vskip -0.7cm
\caption{(a) Value of $N$ that maximizes $\Omega(N)$ in Fig.~\ref{subtract}(a)
as a
function of $T$. (b) Comparison of the time scale of the
maxima from Fig.~\ref{subtract}(a)
with the time scale $\Delta t^*$. }
\label{max_position}
\end{figure}

To understand the differences between the spatially
localized and the randomly distributed dynamic fluctuations,
in Fig.~\ref{subtract}(a) we show the result of subtracting
$\Omega_R(N)$ from $\Omega(N)$.  For $N=1$
particles the result is zero as there is no distinction between
the two calculations.  Likewise for $N \rightarrow 6400$ (the
total number of $A$ particles),
the two calculations are identical.
At intermediate numbers of particles, nonzero values are found
in Fig.~\ref{subtract}(a) indicating nontrivial spatially
localized values of $\Omega$.  In particular, for colder
temperatures the dynamical fluctuations are larger [higher
curves in Fig.~\ref{subtract}(a)] and the subsystem size with
the largest fluctuations grows slightly [peak position shifts
rightward in Fig.~\ref{subtract}(a)].  This last observation is
quantified in Fig.~\ref{max_position} which shows the peak
position of Fig.~\ref{subtract}(a) as a function of $T$.
The peak occurs for larger subsystems
at colder temperatures.  This indicates the size of regions
with maximally variable dynamics contain about 50 particles for
the coldest samples.  For hotter samples, the maximum shifts to
smaller system sizes; for the hottest data, $\Omega$ for compact
subsystems is nearly indistinguishable from $\Omega$ calculated
for random particles, and we cannot identify a maximum.  In turn,
Fig.~\ref{subtract}(b) shows the behavior for the colloidal
suspensions, which is similar to that found for the Lennard-Jones
system. Namely, there is a peak in Fig.~\ref{subtract}(b)
at a specific subsystem size, and the position and height of this peak is
larger for volume fractions closer to the glass transition volume
fraction ($\phi_{\rm g} \approx 0.58$
\cite{weeks00}).  An intriguing difference is that the peak position
indicates that larger subsystems containing a few hundred particles
are maximally heterogeneous, as compared to
Fig.~\ref{subtract}(a) 
which peaks at subsystem sizes containing a few tens of particles.
Earlier work with the Lennard-Jones system found mobile clusters
containing 10-30 particles \cite{donati99pre} or 40 particles
\cite{appignanesi06} at the coldest temperatures, in agreement with
our result.  Likewise, earlier analyses of the same colloidal
data found the largest mobile clusters contained $\sim 50$
particles on average for $\phi=0.562$ \cite{weeks00}.  The new
result seen here is how the sample behaves over larger length
scales.  For example, Fig.~\ref{subtract}(a) shows that there is
still nontrivial spatial heterogeneity for subsystems containing
$N=1000$ particles, more than an order of magnitude larger than
the $N$ corresponding to the peak.  This is strong evidence that
the dynamically heterogeneous regions of size $N \sim 50$ are not randomly
distributed throughout the sample but are themselves spatially
clustered.

In the preceding analysis, $\Omega^2_S$ is averaged over time
scales to highlight the $N$ dependence of the dynamical fluctuations.
  We now turn to the complementary case,
and average  $\Omega^2_S$ over subsystem sizes $N$ to find the time scale $\Delta
t$  of the same fluctuations.  Since the value of $N$ covers several
order of magnitude,
we average $\Omega_S(t',t'')$ 
over subsystems with sizes $N$ picked to be evenly distributed in
$\log(N)$, with $N$ ranging from 1 to the system size.  We first
average over all subsystems of a given size $N$, and then average
$\Omega^2(N,t',t'')$ over $N$ and $(t',t'')$ with fixed time
interval $\Delta t = t'' - t'$ to result in $\Omega(\Delta t)$.
This is shown in Fig.~\ref{max_position}(b) where the peak of each curve is
marked with a red circle.  For comparison, the time scale $\Delta
t^*$ is indicated for each data set with a yellow circle; it appears
the peak of $\Omega(\Delta t)$ is always close to the non-Gaussian time scale
$\Delta t^*$.  As expected, the time scale of maximum dynamical
heterogeneity [as measured by $\Omega(\Delta t)$] grows as the
system approaches the glass transition. 

\section{Conclusions}

In summary, we have constructed a new measure of spatial and
temporal dynamic heterogeneity.  The measure does not require
defining subsets of mobile or immobile particles, but rather
looks for fluctuations away from the large system behavior.
Additionally, it allows us to examine dynamical heterogeneity on
a variety of length scales, showing that the approach to the
large system limit is slower than would be expected for randomly
distributed fluctuations in the dynamics.  The method can be
straightforwardly applied to experimental systems such as the
dense colloidal solution we examine; it does not require
finite-size scaling, for example.  While we focus on particle
motion where the mean square displacement is the null hypothesis,
the method can be generalized to any other spatially and
temporally fluctuating quantities, as long as there is a
well-defined null hypothesis based on the large system limit.

The work of E.R.W was supported by a grant from the
National Science Foundation (DMR-1609763).
GAA and JARF acknowledge suport form CONICET, UNS and
ANPCyT(PICT2015/1893).  Inspiration for this work came from past
conversations with Prof.~Walter Kob.

\bibliography{eric}

\begin{thebibliography}{50}%
\makeatletter
\providecommand \@ifxundefined [1]{%
 \@ifx{#1\undefined}
}%
\providecommand \@ifnum [1]{%
 \ifnum #1\expandafter \@firstoftwo
 \else \expandafter \@secondoftwo
 \fi
}%
\providecommand \@ifx [1]{%
 \ifx #1\expandafter \@firstoftwo
 \else \expandafter \@secondoftwo
 \fi
}%
\providecommand \natexlab [1]{#1}%
\providecommand \enquote  [1]{``#1''}%
\providecommand \bibnamefont  [1]{#1}%
\providecommand \bibfnamefont [1]{#1}%
\providecommand \citenamefont [1]{#1}%
\providecommand \href@noop [0]{\@secondoftwo}%
\providecommand \href [0]{\begingroup \@sanitize@url \@href}%
\providecommand \@href[1]{\@@startlink{#1}\@@href}%
\providecommand \@@href[1]{\endgroup#1\@@endlink}%
\providecommand \@sanitize@url [0]{\catcode `\\12\catcode `\$12\catcode
  `\&12\catcode `\#12\catcode `\^12\catcode `\_12\catcode `\%12\relax}%
\providecommand \@@startlink[1]{}%
\providecommand \@@endlink[0]{}%
\providecommand \url  [0]{\begingroup\@sanitize@url \@url }%
\providecommand \@url [1]{\endgroup\@href {#1}{\urlprefix }}%
\providecommand \urlprefix  [0]{URL }%
\providecommand \Eprint [0]{\href }%
\providecommand \doibase [0]{http://dx.doi.org/}%
\providecommand \selectlanguage [0]{\@gobble}%
\providecommand \bibinfo  [0]{\@secondoftwo}%
\providecommand \bibfield  [0]{\@secondoftwo}%
\providecommand \translation [1]{[#1]}%
\providecommand \BibitemOpen [0]{}%
\providecommand \bibitemStop [0]{}%
\providecommand \bibitemNoStop [0]{.\EOS\space}%
\providecommand \EOS [0]{\spacefactor3000\relax}%
\providecommand \BibitemShut  [1]{\csname bibitem#1\endcsname}%
\let\auto@bib@innerbib\@empty
\bibitem [{\citenamefont {Angell}(1995)}]{angell95}%
  \BibitemOpen
  \bibfield  {author} {\bibinfo {author} {\bibfnamefont {C.~A.}\ \bibnamefont
  {Angell}},\ }\href {\doibase 10.1126/science.267.5206.1924} {\bibfield
  {journal} {\bibinfo  {journal} {Science}\ }\textbf {\bibinfo {volume}
  {267}},\ \bibinfo {pages} {1924} (\bibinfo {year} {1995})}\BibitemShut
  {NoStop}%
\bibitem [{\citenamefont {Angell}\ \emph {et~al.}(2000)\citenamefont {Angell},
  \citenamefont {Ngai}, \citenamefont {McKenna}, \citenamefont {McMillan},\
  and\ \citenamefont {Martin}}]{angell00}%
  \BibitemOpen
  \bibfield  {author} {\bibinfo {author} {\bibfnamefont {C.~A.}\ \bibnamefont
  {Angell}}, \bibinfo {author} {\bibfnamefont {K.~L.}\ \bibnamefont {Ngai}},
  \bibinfo {author} {\bibfnamefont {G.~B.}\ \bibnamefont {McKenna}}, \bibinfo
  {author} {\bibfnamefont {P.~F.}\ \bibnamefont {McMillan}}, \ and\ \bibinfo
  {author} {\bibfnamefont {S.~W.}\ \bibnamefont {Martin}},\ }\href {\doibase
  10.1063/1.1286035} {\bibfield  {journal} {\bibinfo  {journal} {J. App.
  Phys.}\ }\textbf {\bibinfo {volume} {88}},\ \bibinfo {pages} {3113} (\bibinfo
  {year} {2000})}\BibitemShut {NoStop}%
\bibitem [{\citenamefont {Langer}(2014)}]{langer14}%
  \BibitemOpen
  \bibfield  {author} {\bibinfo {author} {\bibfnamefont {J.~S.}\ \bibnamefont
  {Langer}},\ }\href {\doibase 10.1088/0034-4885/77/4/042501} {\bibfield
  {journal} {\bibinfo  {journal} {Rep. Prog. Phys.}\ }\textbf {\bibinfo
  {volume} {77}},\ \bibinfo {pages} {042501} (\bibinfo {year}
  {2014})}\BibitemShut {NoStop}%
\bibitem [{\citenamefont {Chandler}\ and\ \citenamefont
  {Garrahan}(2010)}]{chandler10}%
  \BibitemOpen
  \bibfield  {author} {\bibinfo {author} {\bibfnamefont {D.}~\bibnamefont
  {Chandler}}\ and\ \bibinfo {author} {\bibfnamefont {J.~P.}\ \bibnamefont
  {Garrahan}},\ }\href {\doibase 10.1146/annurev.physchem.040808.090405}
  {\bibfield  {journal} {\bibinfo  {journal} {Ann. Rev. Phys. Chem.}\ }\textbf
  {\bibinfo {volume} {61}},\ \bibinfo {pages} {191} (\bibinfo {year}
  {2010})}\BibitemShut {NoStop}%
\bibitem [{\citenamefont {Biroli}\ and\ \citenamefont
  {Garrahan}(2013)}]{biroli13}%
  \BibitemOpen
  \bibfield  {author} {\bibinfo {author} {\bibfnamefont {G.}~\bibnamefont
  {Biroli}}\ and\ \bibinfo {author} {\bibfnamefont {J.~P.}\ \bibnamefont
  {Garrahan}},\ }\href {\doibase 10.1063/1.4795539} {\bibfield  {journal}
  {\bibinfo  {journal} {J. Chem. Phys.}\ }\textbf {\bibinfo {volume} {138}},\
  \bibinfo {pages} {12A301} (\bibinfo {year} {2013})}\BibitemShut {NoStop}%
\bibitem [{\citenamefont {Ediger}\ and\ \citenamefont
  {Harrowell}(2012)}]{ediger12}%
  \BibitemOpen
  \bibfield  {author} {\bibinfo {author} {\bibfnamefont {M.~D.}\ \bibnamefont
  {Ediger}}\ and\ \bibinfo {author} {\bibfnamefont {P.}~\bibnamefont
  {Harrowell}},\ }\href {\doibase 10.1063/1.4747326} {\bibfield  {journal}
  {\bibinfo  {journal} {J. Chem. Phys.}\ }\textbf {\bibinfo {volume} {137}},\
  \bibinfo {pages} {080901} (\bibinfo {year} {2012})}\BibitemShut {NoStop}%
\bibitem [{\citenamefont {Lubchenko}\ and\ \citenamefont
  {Wolynes}(2007)}]{lubchenko07}%
  \BibitemOpen
  \bibfield  {author} {\bibinfo {author} {\bibfnamefont {V.}~\bibnamefont
  {Lubchenko}}\ and\ \bibinfo {author} {\bibfnamefont {P.~G.}\ \bibnamefont
  {Wolynes}},\ }\href {\doibase 10.1146/annurev.physchem.58.032806.104653}
  {\bibfield  {journal} {\bibinfo  {journal} {Ann. Rev. Phys. Chem.}\ }\textbf
  {\bibinfo {volume} {58}},\ \bibinfo {pages} {235} (\bibinfo {year}
  {2007})}\BibitemShut {NoStop}%
\bibitem [{\citenamefont {Cavagna}(2009)}]{cavagna09}%
  \BibitemOpen
  \bibfield  {author} {\bibinfo {author} {\bibfnamefont {A.}~\bibnamefont
  {Cavagna}},\ }\href {\doibase 10.1016/j.physrep.2009.03.003} {\bibfield
  {journal} {\bibinfo  {journal} {Phys. Rep.}\ }\textbf {\bibinfo {volume}
  {476}},\ \bibinfo {pages} {51} (\bibinfo {year} {2009})}\BibitemShut
  {NoStop}%
\bibitem [{\citenamefont {Sillescu}(1999)}]{sillescu99}%
  \BibitemOpen
  \bibfield  {author} {\bibinfo {author} {\bibfnamefont {H.}~\bibnamefont
  {Sillescu}},\ }\href {\doibase 10.1016/s0022-3093(98)00831-x} {\bibfield
  {journal} {\bibinfo  {journal} {J. Non-Cryst. Solids}\ }\textbf {\bibinfo
  {volume} {243}},\ \bibinfo {pages} {81} (\bibinfo {year} {1999})}\BibitemShut
  {NoStop}%
\bibitem [{\citenamefont {Ediger}(2000)}]{ediger00}%
  \BibitemOpen
  \bibfield  {author} {\bibinfo {author} {\bibfnamefont {M.~D.}\ \bibnamefont
  {Ediger}},\ }\href {\doibase 10.1146/annurev.physchem.51.1.99} {\bibfield
  {journal} {\bibinfo  {journal} {Ann. Rev. Phys. Chem.}\ }\textbf {\bibinfo
  {volume} {51}},\ \bibinfo {pages} {99} (\bibinfo {year} {2000})}\BibitemShut
  {NoStop}%
\bibitem [{\citenamefont {Glotzer}(2000)}]{glotzer00}%
  \BibitemOpen
  \bibfield  {author} {\bibinfo {author} {\bibfnamefont {S.~C.}\ \bibnamefont
  {Glotzer}},\ }\bibfield  {booktitle} {\emph {\bibinfo {booktitle} {Physics of
  Non-Crystalline Solids 9}},\ }\href {\doibase 10.1016/s0022-3093(00)00225-8}
  {\bibfield  {journal} {\bibinfo  {journal} {J. Non-Cryst. Solids}\ }\textbf
  {\bibinfo {volume} {274}},\ \bibinfo {pages} {342} (\bibinfo {year}
  {2000})}\BibitemShut {NoStop}%
\bibitem [{\citenamefont {Hempel}\ \emph {et~al.}(2000)\citenamefont {Hempel},
  \citenamefont {Hempel}, \citenamefont {Hensel}, \citenamefont {Schick},\ and\
  \citenamefont {Donth}}]{hempel00}%
  \BibitemOpen
  \bibfield  {author} {\bibinfo {author} {\bibfnamefont {E.}~\bibnamefont
  {Hempel}}, \bibinfo {author} {\bibfnamefont {G.}~\bibnamefont {Hempel}},
  \bibinfo {author} {\bibfnamefont {A.}~\bibnamefont {Hensel}}, \bibinfo
  {author} {\bibfnamefont {C.}~\bibnamefont {Schick}}, \ and\ \bibinfo {author}
  {\bibfnamefont {E.}~\bibnamefont {Donth}},\ }\href {\doibase
  10.1021/jp991153f} {\bibfield  {journal} {\bibinfo  {journal} {J. Phys. Chem.
  B}\ }\textbf {\bibinfo {volume} {104}},\ \bibinfo {pages} {2460} (\bibinfo
  {year} {2000})}\BibitemShut {NoStop}%
\bibitem [{\citenamefont {Richert}(2002)}]{richert02}%
  \BibitemOpen
  \bibfield  {author} {\bibinfo {author} {\bibfnamefont {R.}~\bibnamefont
  {Richert}},\ }\href {\doibase 10.1088/0953-8984/14/23/201} {\bibfield
  {journal} {\bibinfo  {journal} {J. Phys.: Condens. Matter}\ }\textbf
  {\bibinfo {volume} {14}},\ \bibinfo {pages} {R703} (\bibinfo {year}
  {2002})}\BibitemShut {NoStop}%
\bibitem [{\citenamefont {Kob}\ \emph {et~al.}(1997)\citenamefont {Kob},
  \citenamefont {Donati}, \citenamefont {Plimpton}, \citenamefont {Poole},\
  and\ \citenamefont {Glotzer}}]{kob97}%
  \BibitemOpen
  \bibfield  {author} {\bibinfo {author} {\bibfnamefont {W.}~\bibnamefont
  {Kob}}, \bibinfo {author} {\bibfnamefont {C.}~\bibnamefont {Donati}},
  \bibinfo {author} {\bibfnamefont {S.~J.}\ \bibnamefont {Plimpton}}, \bibinfo
  {author} {\bibfnamefont {P.~H.}\ \bibnamefont {Poole}}, \ and\ \bibinfo
  {author} {\bibfnamefont {S.~C.}\ \bibnamefont {Glotzer}},\ }\href {\doibase
  10.1103/physrevlett.79.2827} {\bibfield  {journal} {\bibinfo  {journal}
  {Phys. Rev. Lett.}\ }\textbf {\bibinfo {volume} {79}},\ \bibinfo {pages}
  {2827} (\bibinfo {year} {1997})}\BibitemShut {NoStop}%
\bibitem [{\citenamefont {Donati}\ \emph {et~al.}(1998)\citenamefont {Donati},
  \citenamefont {Douglas}, \citenamefont {Kob}, \citenamefont {Plimpton},
  \citenamefont {Poole},\ and\ \citenamefont {Glotzer}}]{donati98}%
  \BibitemOpen
  \bibfield  {author} {\bibinfo {author} {\bibfnamefont {C.}~\bibnamefont
  {Donati}}, \bibinfo {author} {\bibfnamefont {J.~F.}\ \bibnamefont {Douglas}},
  \bibinfo {author} {\bibfnamefont {W.}~\bibnamefont {Kob}}, \bibinfo {author}
  {\bibfnamefont {S.~J.}\ \bibnamefont {Plimpton}}, \bibinfo {author}
  {\bibfnamefont {P.~H.}\ \bibnamefont {Poole}}, \ and\ \bibinfo {author}
  {\bibfnamefont {S.~C.}\ \bibnamefont {Glotzer}},\ }\href {\doibase
  10.1103/physrevlett.80.2338} {\bibfield  {journal} {\bibinfo  {journal}
  {Phys. Rev. Lett.}\ }\textbf {\bibinfo {volume} {80}},\ \bibinfo {pages}
  {2338} (\bibinfo {year} {1998})}\BibitemShut {NoStop}%
\bibitem [{\citenamefont {Donati}\ \emph
  {et~al.}(1999{\natexlab{a}})\citenamefont {Donati}, \citenamefont {Glotzer},\
  and\ \citenamefont {Poole}}]{donati99}%
  \BibitemOpen
  \bibfield  {author} {\bibinfo {author} {\bibfnamefont {C.}~\bibnamefont
  {Donati}}, \bibinfo {author} {\bibfnamefont {S.~C.}\ \bibnamefont {Glotzer}},
  \ and\ \bibinfo {author} {\bibfnamefont {P.~H.}\ \bibnamefont {Poole}},\
  }\href {\doibase 10.1103/physrevlett.82.5064} {\bibfield  {journal} {\bibinfo
   {journal} {Phys. Rev. Lett.}\ }\textbf {\bibinfo {volume} {82}},\ \bibinfo
  {pages} {5064} (\bibinfo {year} {1999}{\natexlab{a}})}\BibitemShut {NoStop}%
\bibitem [{\citenamefont {Garrahan}\ and\ \citenamefont
  {Chandler}(2003)}]{garrahan03}%
  \BibitemOpen
  \bibfield  {author} {\bibinfo {author} {\bibfnamefont {J.~P.}\ \bibnamefont
  {Garrahan}}\ and\ \bibinfo {author} {\bibfnamefont {D.}~\bibnamefont
  {Chandler}},\ }\href {\doibase 10.1073/pnas.1233719100} {\bibfield  {journal}
  {\bibinfo  {journal} {Proc. Nat. Acad. Sci.}\ }\textbf {\bibinfo {volume}
  {100}},\ \bibinfo {pages} {9710} (\bibinfo {year} {2003})}\BibitemShut
  {NoStop}%
\bibitem [{\citenamefont {Keys}\ \emph {et~al.}(2011)\citenamefont {Keys},
  \citenamefont {Hedges}, \citenamefont {Garrahan}, \citenamefont {Glotzer},\
  and\ \citenamefont {Chandler}}]{keys11}%
  \BibitemOpen
  \bibfield  {author} {\bibinfo {author} {\bibfnamefont {A.~S.}\ \bibnamefont
  {Keys}}, \bibinfo {author} {\bibfnamefont {L.~O.}\ \bibnamefont {Hedges}},
  \bibinfo {author} {\bibfnamefont {J.~P.}\ \bibnamefont {Garrahan}}, \bibinfo
  {author} {\bibfnamefont {S.~C.}\ \bibnamefont {Glotzer}}, \ and\ \bibinfo
  {author} {\bibfnamefont {D.}~\bibnamefont {Chandler}},\ }\href {\doibase
  10.1103/PhysRevX.1.021013} {\bibfield  {journal} {\bibinfo  {journal} {Phys.
  Rev. X}\ }\textbf {\bibinfo {volume} {1}},\ \bibinfo {pages} {021013}
  (\bibinfo {year} {2011})}\BibitemShut {NoStop}%
\bibitem [{\citenamefont {Hurley}\ and\ \citenamefont
  {Harrowell}(1995)}]{hurley95}%
  \BibitemOpen
  \bibfield  {author} {\bibinfo {author} {\bibfnamefont {M.~M.}\ \bibnamefont
  {Hurley}}\ and\ \bibinfo {author} {\bibfnamefont {P.}~\bibnamefont
  {Harrowell}},\ }\href {\doibase 10.1103/physreve.52.1694} {\bibfield
  {journal} {\bibinfo  {journal} {Phys. Rev. E}\ }\textbf {\bibinfo {volume}
  {52}},\ \bibinfo {pages} {1694} (\bibinfo {year} {1995})}\BibitemShut
  {NoStop}%
\bibitem [{\citenamefont {Hurley}\ and\ \citenamefont
  {Harrowell}(1996)}]{hurley96}%
  \BibitemOpen
  \bibfield  {author} {\bibinfo {author} {\bibfnamefont {M.~M.}\ \bibnamefont
  {Hurley}}\ and\ \bibinfo {author} {\bibfnamefont {P.}~\bibnamefont
  {Harrowell}},\ }\href@noop {} {\bibfield  {journal} {\bibinfo  {journal} {J.
  Chem. Phys.}\ }\textbf {\bibinfo {volume} {105}},\ \bibinfo {pages} {10521}
  (\bibinfo {year} {1996})}\BibitemShut {NoStop}%
\bibitem [{\citenamefont {Marcus}\ \emph {et~al.}(1999)\citenamefont {Marcus},
  \citenamefont {Schofield},\ and\ \citenamefont {Rice}}]{marcus99}%
  \BibitemOpen
  \bibfield  {author} {\bibinfo {author} {\bibfnamefont {A.~H.}\ \bibnamefont
  {Marcus}}, \bibinfo {author} {\bibfnamefont {J.}~\bibnamefont {Schofield}}, \
  and\ \bibinfo {author} {\bibfnamefont {S.~A.}\ \bibnamefont {Rice}},\ }\href
  {\doibase 10.1103/physreve.60.5725} {\bibfield  {journal} {\bibinfo
  {journal} {Phys. Rev. E}\ }\textbf {\bibinfo {volume} {60}},\ \bibinfo
  {pages} {5725} (\bibinfo {year} {1999})}\BibitemShut {NoStop}%
\bibitem [{\citenamefont {Rahman}(1964)}]{rahman64}%
  \BibitemOpen
  \bibfield  {author} {\bibinfo {author} {\bibfnamefont {A.}~\bibnamefont
  {Rahman}},\ }\href {\doibase 10.1103/physrev.136.a405} {\bibfield  {journal}
  {\bibinfo  {journal} {Phys. Rev.}\ }\textbf {\bibinfo {volume} {136}},\
  \bibinfo {pages} {A405} (\bibinfo {year} {1964})}\BibitemShut {NoStop}%
\bibitem [{\citenamefont {Flenner}\ and\ \citenamefont
  {Szamel}(2007)}]{flenner07}%
  \BibitemOpen
  \bibfield  {author} {\bibinfo {author} {\bibfnamefont {E.}~\bibnamefont
  {Flenner}}\ and\ \bibinfo {author} {\bibfnamefont {G.}~\bibnamefont
  {Szamel}},\ }\href {\doibase 10.1088/0953-8984/19/20/205125} {\bibfield
  {journal} {\bibinfo  {journal} {J. Phys.: Condens. Matter}\ }\textbf
  {\bibinfo {volume} {19}},\ \bibinfo {pages} {205125} (\bibinfo {year}
  {2007})}\BibitemShut {NoStop}%
\bibitem [{\citenamefont {Doliwa}\ and\ \citenamefont
  {Heuer}(2000)}]{doliwa00}%
  \BibitemOpen
  \bibfield  {author} {\bibinfo {author} {\bibfnamefont {B.}~\bibnamefont
  {Doliwa}}\ and\ \bibinfo {author} {\bibfnamefont {A.}~\bibnamefont {Heuer}},\
  }\href {\doibase 10.1103/physreve.61.6898} {\bibfield  {journal} {\bibinfo
  {journal} {Phys. Rev. E}\ }\textbf {\bibinfo {volume} {61}},\ \bibinfo
  {pages} {6898} (\bibinfo {year} {2000})}\BibitemShut {NoStop}%
\bibitem [{\citenamefont {Weeks}\ \emph {et~al.}(2007)\citenamefont {Weeks},
  \citenamefont {Crocker},\ and\ \citenamefont {Weitz}}]{weeks07cor}%
  \BibitemOpen
  \bibfield  {author} {\bibinfo {author} {\bibfnamefont {E.~R.}\ \bibnamefont
  {Weeks}}, \bibinfo {author} {\bibfnamefont {J.~C.}\ \bibnamefont {Crocker}},
  \ and\ \bibinfo {author} {\bibfnamefont {D.~A.}\ \bibnamefont {Weitz}},\
  }\href {\doibase 10.1088/0953-8984/19/20/205131} {\bibfield  {journal}
  {\bibinfo  {journal} {J. Phys.: Condens. Matter}\ }\textbf {\bibinfo {volume}
  {19}},\ \bibinfo {pages} {205131} (\bibinfo {year} {2007})}\BibitemShut
  {NoStop}%
\bibitem [{\citenamefont {La\v{c}evi\'{c}}\ \emph {et~al.}(2003)\citenamefont
  {La\v{c}evi\'{c}}, \citenamefont {Starr}, \citenamefont {Schr{\o}der},\ and\
  \citenamefont {Glotzer}}]{lacevic03}%
  \BibitemOpen
  \bibfield  {author} {\bibinfo {author} {\bibfnamefont {N.}~\bibnamefont
  {La\v{c}evi\'{c}}}, \bibinfo {author} {\bibfnamefont {F.~W.}\ \bibnamefont
  {Starr}}, \bibinfo {author} {\bibfnamefont {T.~B.}\ \bibnamefont
  {Schr{\o}der}}, \ and\ \bibinfo {author} {\bibfnamefont {S.~C.}\ \bibnamefont
  {Glotzer}},\ }\href@noop {} {\bibfield  {journal} {\bibinfo  {journal} {J.
  Chem. Phys.}\ }\textbf {\bibinfo {volume} {119}},\ \bibinfo {pages} {7372}
  (\bibinfo {year} {2003})}\BibitemShut {NoStop}%
\bibitem [{\citenamefont {Keys}\ \emph {et~al.}(2007)\citenamefont {Keys},
  \citenamefont {Abate}, \citenamefont {Glotzer},\ and\ \citenamefont
  {Durian}}]{keys07}%
  \BibitemOpen
  \bibfield  {author} {\bibinfo {author} {\bibfnamefont {A.~S.}\ \bibnamefont
  {Keys}}, \bibinfo {author} {\bibfnamefont {A.~R.}\ \bibnamefont {Abate}},
  \bibinfo {author} {\bibfnamefont {S.~C.}\ \bibnamefont {Glotzer}}, \ and\
  \bibinfo {author} {\bibfnamefont {D.~J.}\ \bibnamefont {Durian}},\ }\href
  {\doibase 10.1038/nphys572} {\bibfield  {journal} {\bibinfo  {journal}
  {Nature Phys.}\ }\textbf {\bibinfo {volume} {3}},\ \bibinfo {pages} {260}
  (\bibinfo {year} {2007})}\BibitemShut {NoStop}%
\bibitem [{\citenamefont {Appignanesi}\ \emph {et~al.}(2006)\citenamefont
  {Appignanesi}, \citenamefont {Rodriguez~Fris}, \citenamefont {Montani},\ and\
  \citenamefont {Kob}}]{appignanesi06}%
  \BibitemOpen
  \bibfield  {author} {\bibinfo {author} {\bibfnamefont {G.~A.}\ \bibnamefont
  {Appignanesi}}, \bibinfo {author} {\bibfnamefont {J.~A.}\ \bibnamefont
  {Rodriguez~Fris}}, \bibinfo {author} {\bibfnamefont {R.~A.}\ \bibnamefont
  {Montani}}, \ and\ \bibinfo {author} {\bibfnamefont {W.}~\bibnamefont
  {Kob}},\ }\href {\doibase 10.1103/physrevlett.96.057801} {\bibfield
  {journal} {\bibinfo  {journal} {Phys. Rev. Lett.}\ }\textbf {\bibinfo
  {volume} {96}},\ \bibinfo {pages} {057801} (\bibinfo {year}
  {2006})}\BibitemShut {NoStop}%
\bibitem [{\citenamefont {Appignanesi}\ and\ \citenamefont
  {Rodriguez~Fris}(2009)}]{appignanesi09}%
  \BibitemOpen
  \bibfield  {author} {\bibinfo {author} {\bibfnamefont {G.~A.}\ \bibnamefont
  {Appignanesi}}\ and\ \bibinfo {author} {\bibfnamefont {J.~A.}\ \bibnamefont
  {Rodriguez~Fris}},\ }\href {\doibase 10.1088/0953-8984/21/20/203103}
  {\bibfield  {journal} {\bibinfo  {journal} {J. Phys.: Condens. Matter}\
  }\textbf {\bibinfo {volume} {21}},\ \bibinfo {pages} {203103} (\bibinfo
  {year} {2009})}\BibitemShut {NoStop}%
\bibitem [{\citenamefont {Kob}\ and\ \citenamefont {Andersen}(1995)}]{kob95a}%
  \BibitemOpen
  \bibfield  {author} {\bibinfo {author} {\bibfnamefont {W.}~\bibnamefont
  {Kob}}\ and\ \bibinfo {author} {\bibfnamefont {H.~C.}\ \bibnamefont
  {Andersen}},\ }\href {\doibase 10.1103/physreve.51.4626} {\bibfield
  {journal} {\bibinfo  {journal} {Phys. Rev. E}\ }\textbf {\bibinfo {volume}
  {51}},\ \bibinfo {pages} {4626} (\bibinfo {year} {1995})}\BibitemShut
  {NoStop}%
\bibitem [{\citenamefont {Weeks}\ \emph {et~al.}(2000)\citenamefont {Weeks},
  \citenamefont {Crocker}, \citenamefont {Levitt}, \citenamefont {Schofield},\
  and\ \citenamefont {Weitz}}]{weeks00}%
  \BibitemOpen
  \bibfield  {author} {\bibinfo {author} {\bibfnamefont {E.~R.}\ \bibnamefont
  {Weeks}}, \bibinfo {author} {\bibfnamefont {J.~C.}\ \bibnamefont {Crocker}},
  \bibinfo {author} {\bibfnamefont {A.~C.}\ \bibnamefont {Levitt}}, \bibinfo
  {author} {\bibfnamefont {A.}~\bibnamefont {Schofield}}, \ and\ \bibinfo
  {author} {\bibfnamefont {D.~A.}\ \bibnamefont {Weitz}},\ }\href {\doibase
  10.1126/science.287.5453.627} {\bibfield  {journal} {\bibinfo  {journal}
  {Science}\ }\textbf {\bibinfo {volume} {287}},\ \bibinfo {pages} {627}
  (\bibinfo {year} {2000})}\BibitemShut {NoStop}%
\bibitem [{\citenamefont {Royall}\ and\ \citenamefont
  {Williams}(2015)}]{royall15}%
  \BibitemOpen
  \bibfield  {author} {\bibinfo {author} {\bibfnamefont {C.~P.}\ \bibnamefont
  {Royall}}\ and\ \bibinfo {author} {\bibfnamefont {S.~R.}\ \bibnamefont
  {Williams}},\ }\href {\doibase 10.1016/j.physrep.2014.11.004} {\bibfield
  {journal} {\bibinfo  {journal} {Phys. Rep.}\ }\textbf {\bibinfo {volume}
  {560}},\ \bibinfo {pages} {1} (\bibinfo {year} {2015})}\BibitemShut {NoStop}%
\bibitem [{\citenamefont {Royall}\ \emph {et~al.}(2008)\citenamefont {Royall},
  \citenamefont {Williams}, \citenamefont {Ohtsuka},\ and\ \citenamefont
  {Tanaka}}]{royall08}%
  \BibitemOpen
  \bibfield  {author} {\bibinfo {author} {\bibfnamefont {C.~P.}\ \bibnamefont
  {Royall}}, \bibinfo {author} {\bibfnamefont {S.~R.}\ \bibnamefont
  {Williams}}, \bibinfo {author} {\bibfnamefont {T.}~\bibnamefont {Ohtsuka}}, \
  and\ \bibinfo {author} {\bibfnamefont {H.}~\bibnamefont {Tanaka}},\ }\href
  {\doibase 10.1038/nmat2219} {\bibfield  {journal} {\bibinfo  {journal} {Nat
  Mater}\ }\textbf {\bibinfo {volume} {7}},\ \bibinfo {pages} {556} (\bibinfo
  {year} {2008})}\BibitemShut {NoStop}%
\bibitem [{\citenamefont {{Lennard-Jones}}(1924)}]{lennardjones24}%
  \BibitemOpen
  \bibfield  {author} {\bibinfo {author} {\bibfnamefont {J.~E.}\ \bibnamefont
  {{Lennard-Jones}}},\ }\href {\doibase 10.1098/rspa.1924.0082} {\bibfield
  {journal} {\bibinfo  {journal} {Proc. Roy. Soc. London A: Mathematical,
  Physical and Engineering Sciences}\ }\textbf {\bibinfo {volume} {106}},\
  \bibinfo {pages} {463} (\bibinfo {year} {1924})}\BibitemShut {NoStop}%
\bibitem [{\citenamefont {Hunter}\ and\ \citenamefont
  {Weeks}(2012)}]{hunter12rpp}%
  \BibitemOpen
  \bibfield  {author} {\bibinfo {author} {\bibfnamefont {G.~L.}\ \bibnamefont
  {Hunter}}\ and\ \bibinfo {author} {\bibfnamefont {E.~R.}\ \bibnamefont
  {Weeks}},\ }\href {\doibase 10.1088/0034-4885/75/6/066501} {\bibfield
  {journal} {\bibinfo  {journal} {Rep. Prog. Phys.}\ }\textbf {\bibinfo
  {volume} {75}},\ \bibinfo {pages} {066501} (\bibinfo {year}
  {2012})}\BibitemShut {NoStop}%
\bibitem [{\citenamefont {Marshall}\ and\ \citenamefont
  {Zukoski}(1990)}]{marshall90}%
  \BibitemOpen
  \bibfield  {author} {\bibinfo {author} {\bibfnamefont {L.}~\bibnamefont
  {Marshall}}\ and\ \bibinfo {author} {\bibfnamefont {C.~F.}\ \bibnamefont
  {Zukoski}},\ }\href {\doibase 10.1021/j100366a030} {\bibfield  {journal}
  {\bibinfo  {journal} {J. Phys. Chem.}\ }\textbf {\bibinfo {volume} {94}},\
  \bibinfo {pages} {1164} (\bibinfo {year} {1990})}\BibitemShut {NoStop}%
\bibitem [{\citenamefont {Pusey}\ and\ \citenamefont {van
  Megen}(1987)}]{pusey87}%
  \BibitemOpen
  \bibfield  {author} {\bibinfo {author} {\bibfnamefont {P.~N.}\ \bibnamefont
  {Pusey}}\ and\ \bibinfo {author} {\bibfnamefont {W.}~\bibnamefont {van
  Megen}},\ }\href {\doibase 10.1103/physrevlett.59.2083} {\bibfield  {journal}
  {\bibinfo  {journal} {Phys. Rev. Lett.}\ }\textbf {\bibinfo {volume} {59}},\
  \bibinfo {pages} {2083} (\bibinfo {year} {1987})}\BibitemShut {NoStop}%
\bibitem [{\citenamefont {Pusey}\ and\ \citenamefont {van
  Megen}(1986)}]{pusey86}%
  \BibitemOpen
  \bibfield  {author} {\bibinfo {author} {\bibfnamefont {P.~N.}\ \bibnamefont
  {Pusey}}\ and\ \bibinfo {author} {\bibfnamefont {W.}~\bibnamefont {van
  Megen}},\ }\href {\doibase 10.1038/320340a0} {\bibfield  {journal} {\bibinfo
  {journal} {Nature}\ }\textbf {\bibinfo {volume} {320}},\ \bibinfo {pages}
  {340} (\bibinfo {year} {1986})}\BibitemShut {NoStop}%
\bibitem [{\citenamefont {H{\"a}rtl}(2001)}]{hartl01}%
  \BibitemOpen
  \bibfield  {author} {\bibinfo {author} {\bibfnamefont {W.}~\bibnamefont
  {H{\"a}rtl}},\ }\href {\doibase 10.1016/s1359-0294(01)00120-0} {\bibfield
  {journal} {\bibinfo  {journal} {Curr. Op. Coll. Int. Sci.}\ }\textbf
  {\bibinfo {volume} {6}},\ \bibinfo {pages} {479} (\bibinfo {year}
  {2001})}\BibitemShut {NoStop}%
\bibitem [{\citenamefont {van Megen}\ and\ \citenamefont
  {Pusey}(1991)}]{vanmegen91}%
  \BibitemOpen
  \bibfield  {author} {\bibinfo {author} {\bibfnamefont {W.}~\bibnamefont {van
  Megen}}\ and\ \bibinfo {author} {\bibfnamefont {P.~N.}\ \bibnamefont
  {Pusey}},\ }\href {\doibase 10.1103/physreva.43.5429} {\bibfield  {journal}
  {\bibinfo  {journal} {Phys. Rev. A}\ }\textbf {\bibinfo {volume} {43}},\
  \bibinfo {pages} {5429} (\bibinfo {year} {1991})}\BibitemShut {NoStop}%
\bibitem [{\citenamefont {Bartsch}\ \emph {et~al.}(1993)\citenamefont
  {Bartsch}, \citenamefont {Frenz}, \citenamefont {Moller},\ and\ \citenamefont
  {Sillescu}}]{bartsch93}%
  \BibitemOpen
  \bibfield  {author} {\bibinfo {author} {\bibfnamefont {E.}~\bibnamefont
  {Bartsch}}, \bibinfo {author} {\bibfnamefont {V.}~\bibnamefont {Frenz}},
  \bibinfo {author} {\bibfnamefont {S.}~\bibnamefont {Moller}}, \ and\ \bibinfo
  {author} {\bibfnamefont {H.}~\bibnamefont {Sillescu}},\ }\href {\doibase
  10.1016/0378-4371(93)90433-5} {\bibfield  {journal} {\bibinfo  {journal}
  {Physica A}\ }\textbf {\bibinfo {volume} {201}},\ \bibinfo {pages} {363}
  (\bibinfo {year} {1993})}\BibitemShut {NoStop}%
\bibitem [{\citenamefont {Meller}\ and\ \citenamefont
  {Stavans}(1992)}]{meller92}%
  \BibitemOpen
  \bibfield  {author} {\bibinfo {author} {\bibfnamefont {A.}~\bibnamefont
  {Meller}}\ and\ \bibinfo {author} {\bibfnamefont {J.}~\bibnamefont
  {Stavans}},\ }\href {\doibase 10.1103/physrevlett.68.3646} {\bibfield
  {journal} {\bibinfo  {journal} {Phys. Rev. Lett.}\ }\textbf {\bibinfo
  {volume} {68}},\ \bibinfo {pages} {3646} (\bibinfo {year}
  {1992})}\BibitemShut {NoStop}%
\bibitem [{\citenamefont {Lindsay}\ and\ \citenamefont
  {Chaikin}(1982)}]{lindsay82}%
  \BibitemOpen
  \bibfield  {author} {\bibinfo {author} {\bibfnamefont {H.~M.}\ \bibnamefont
  {Lindsay}}\ and\ \bibinfo {author} {\bibfnamefont {P.~M.}\ \bibnamefont
  {Chaikin}},\ }\href {\doibase 10.1063/1.443417} {\bibfield  {journal}
  {\bibinfo  {journal} {J. Chem. Phys.}\ }\textbf {\bibinfo {volume} {76}},\
  \bibinfo {pages} {3774} (\bibinfo {year} {1982})}\BibitemShut {NoStop}%
\bibitem [{\citenamefont {Kurita}\ \emph {et~al.}(2012)\citenamefont {Kurita},
  \citenamefont {Ruffner},\ and\ \citenamefont {Weeks}}]{kurita12}%
  \BibitemOpen
  \bibfield  {author} {\bibinfo {author} {\bibfnamefont {R.}~\bibnamefont
  {Kurita}}, \bibinfo {author} {\bibfnamefont {D.~B.}\ \bibnamefont {Ruffner}},
  \ and\ \bibinfo {author} {\bibfnamefont {E.~R.}\ \bibnamefont {Weeks}},\
  }\href {\doibase 10.1038/ncomms2114} {\bibfield  {journal} {\bibinfo
  {journal} {Nature Comm.}\ }\textbf {\bibinfo {volume} {3}},\ \bibinfo {pages}
  {1127} (\bibinfo {year} {2012})}\BibitemShut {NoStop}%
\bibitem [{\citenamefont {Dinsmore}\ \emph {et~al.}(2001)\citenamefont
  {Dinsmore}, \citenamefont {Weeks}, \citenamefont {Prasad}, \citenamefont
  {Levitt},\ and\ \citenamefont {Weitz}}]{dinsmore01}%
  \BibitemOpen
  \bibfield  {author} {\bibinfo {author} {\bibfnamefont {A.~D.}\ \bibnamefont
  {Dinsmore}}, \bibinfo {author} {\bibfnamefont {E.~R.}\ \bibnamefont {Weeks}},
  \bibinfo {author} {\bibfnamefont {V.}~\bibnamefont {Prasad}}, \bibinfo
  {author} {\bibfnamefont {A.~C.}\ \bibnamefont {Levitt}}, \ and\ \bibinfo
  {author} {\bibfnamefont {D.~A.}\ \bibnamefont {Weitz}},\ }\href {\doibase
  10.1364/AO.40.004152} {\bibfield  {journal} {\bibinfo  {journal} {App.
  Optics}\ }\textbf {\bibinfo {volume} {40}},\ \bibinfo {pages} {4152}
  (\bibinfo {year} {2001})}\BibitemShut {NoStop}%
\bibitem [{\citenamefont {Crocker}\ and\ \citenamefont
  {Grier}(1996)}]{crocker96}%
  \BibitemOpen
  \bibfield  {author} {\bibinfo {author} {\bibfnamefont {J.~C.}\ \bibnamefont
  {Crocker}}\ and\ \bibinfo {author} {\bibfnamefont {D.~G.}\ \bibnamefont
  {Grier}},\ }\href {\doibase 10.1006/jcis.1996.0217} {\bibfield  {journal}
  {\bibinfo  {journal} {J. Colloid Interface Sci.}\ }\textbf {\bibinfo {volume}
  {179}},\ \bibinfo {pages} {298} (\bibinfo {year} {1996})}\BibitemShut
  {NoStop}%
\bibitem [{\citenamefont {Ohmine}\ and\ \citenamefont
  {Tanaka}(1993)}]{ohmine1993fluctuation}%
  \BibitemOpen
  \bibfield  {author} {\bibinfo {author} {\bibfnamefont {I.}~\bibnamefont
  {Ohmine}}\ and\ \bibinfo {author} {\bibfnamefont {H.}~\bibnamefont
  {Tanaka}},\ }\href@noop {} {\bibfield  {journal} {\bibinfo  {journal} {Chem.
  Reviews}\ }\textbf {\bibinfo {volume} {93}},\ \bibinfo {pages} {2545}
  (\bibinfo {year} {1993})}\BibitemShut {NoStop}%
\bibitem [{\citenamefont {La~Nave}\ and\ \citenamefont
  {Sciortino}(2004)}]{la2004static}%
  \BibitemOpen
  \bibfield  {author} {\bibinfo {author} {\bibfnamefont {E.}~\bibnamefont
  {La~Nave}}\ and\ \bibinfo {author} {\bibfnamefont {F.}~\bibnamefont
  {Sciortino}},\ }\href@noop {} {\bibfield  {journal} {\bibinfo  {journal} {J.
  Phys. Chem. B}\ }\textbf {\bibinfo {volume} {108}},\ \bibinfo {pages} {19663}
  (\bibinfo {year} {2004})}\BibitemShut {NoStop}%
\bibitem [{\citenamefont {Chandler}(1987)}]{chandlerbook}%
  \BibitemOpen
  \bibfield  {author} {\bibinfo {author} {\bibfnamefont {D.}~\bibnamefont
  {Chandler}},\ }\href@noop {} {\bibfield  {journal} {\bibinfo  {journal}
  {Introduction to Modern Statistical Mechanics, by David Chandler, pp. 288.
  Oxford University Press, Sep 1987.}\ ,\ \bibinfo {pages} {288}} (\bibinfo
  {year} {1987})}\BibitemShut {NoStop}%
\bibitem [{\citenamefont {Donati}\ \emph
  {et~al.}(1999{\natexlab{b}})\citenamefont {Donati}, \citenamefont {Glotzer},
  \citenamefont {Poole}, \citenamefont {Kob},\ and\ \citenamefont
  {Plimpton}}]{donati99pre}%
  \BibitemOpen
  \bibfield  {author} {\bibinfo {author} {\bibfnamefont {C.}~\bibnamefont
  {Donati}}, \bibinfo {author} {\bibfnamefont {S.~C.}\ \bibnamefont {Glotzer}},
  \bibinfo {author} {\bibfnamefont {P.~H.}\ \bibnamefont {Poole}}, \bibinfo
  {author} {\bibfnamefont {W.}~\bibnamefont {Kob}}, \ and\ \bibinfo {author}
  {\bibfnamefont {S.~J.}\ \bibnamefont {Plimpton}},\ }\href {\doibase
  10.1103/physreve.60.3107} {\bibfield  {journal} {\bibinfo  {journal} {Phys.
  Rev. E}\ }\textbf {\bibinfo {volume} {60}},\ \bibinfo {pages} {3107}
  (\bibinfo {year} {1999}{\natexlab{b}})}\BibitemShut {NoStop}%
\end{thebibliography}%

\end{document}